\documentclass{article}
\usepackage[english]{babel}
\usepackage{ismir,amsmath,cite,url}
\usepackage{graphicx}
\usepackage{color}
\usepackage{subcaption}
\usepackage{booktabs}
\usepackage{subcaption}

\twoauthors
   {Dorien Herremans} {Singapore University of Technology and Design \\ Information Systems, Technology and Design Pillar \\\ \& Institute of High Performance Computing \\ A*STAR, Singapore \\ {\tt dorien\_herremans@sutd.edu.sg}}
  {Tom Bergmans} {University of Antwerp \\ Department of Engineering Management}

\title{Hit song prediction based on early adopter data and audio features}

\let\OLDthebibliography\thebibliography
\renewcommand\thebibliography[1]{
  \OLDthebibliography{#1}
  \setlength{\parskip}{0pt}
  \setlength{\itemsep}{0pt plus 0.3ex}
}

\date{\today}

\begin{document}
\maketitle

\begin{abstract}

Billions of USD are invested in new artists and songs by the music industry every year. This research provides a new strategy for assessing the hit potential of songs, which can help record companies support their investment decisions. A number of models were developed that use both audio data, and a novel feature based on social media listening behaviour. The results show that models based on early adopter behaviour perform well when predicting top~20 dance hits. 

\end{abstract}

\vspace{-.4cm}
\section{Introduction}

\emph{``I've written about 78 top 10 songs, and I still don't know what a hit song is." Lamont Dozier, one of the most successful producers behind the Motown label~\cite{lindvall2011song}}

Given the huge investments in new talent by the music industry~\cite{ifpi2012}, the question of predicting hits is extremely relevant. 
The science of hit song prediction has had a controversial history, as early studies such as \cite{pachet2008hit, borg2011makes} showed that random oracles can not always be outperformed when it comes to predicting hits. More recent studies, however, have managed to developed more accurate classifiers for separating hits from non-hits \cite{herremans2014dance, ni2011hit}. These successful hit prediction models use various audio features, including temporal features~\cite{herremans2014dance}, or lyrics\cite{dhanaraj2005automatic}. In this study, we not only focus on audio features, but also include social media listening behaviours to identify early adopters.

%If we could define a set of rules or characteristics of what differentiates hit songs from music that never reaches the big crowd, musicians or producers could take these things in mind and use it as guidelines. In other words: What makes a song a full blown hit and how can we predict the popularity of a musician or a song. To answer this million dollar question we can rely on the most recent machine learning techniques and define patterns in vast amounts of data to discover the building blocks of the hit-formula.

\vspace{-.2cm}

\section{Early adopters}

\emph{‘There are people \dots with some sort of sixth sense for hit songs, such that they listen to hit songs even before they actually climb to the top of the record charts.’} \cite{smit2013hit}

In \cite{rogers2010diffusion}, Rogers defines early adopters as individuals with `the highest degree of opinion leadership among the other adopter categories'. This group of adopters is typically younger and has a higher social status and is more `socially forward' than others. 

%Realize judicious choice of adoption will help them maintain central communication position'. 

The idea of early adopters fits in the context of today's music listening experience and the way hit songs are spread among peers on platforms such as Last.FM. Inspired by preliminary research of \cite{smit2013hit}, the models in this research will use listening behaviour and hit listings information to identify a group of early adopters and embed this knowledge in a model for future hit predictions. In contrast to the original thesis by \cite{smit2013hit}, which only uses 5 songs, we will use a much larger dataset in order to draw more general conclusions. The creation of this dataset is described in the next section.

\vspace{-.4cm}
\section{Dataset}

\subsection{Hits versus non-hits}
A list of hit songs was parsed from dance charts listed on the Belgian website `The Ultratop 50'\footnote{\url{www.ultratop.be/nl/ultratop50}}. The top 20 songs were used to identify `hits'. In order to obtain a list of non-hits, previous research has used low ranking songs from hit lists (e.g. top 30-50 in \cite{herremans2014dance}). In this research, however, we were able to use `The Bubbling Under chart', also from the Ultratop 50 site, which contains a list of 20 upcoming songs identified by music experts. We filtered out any song that eventually hit the charts, and only kept those that never became a hit. The data from 2/07/11 until 16/11/13 was recorded our the dataset, resulting in a total of 8,750 songs, of which 982 were unique.

\subsection{Listening behaviour}

\begin{figure*}[h!] \centering
\begin{subfigure}[h]{0.32\textwidth} \centering
\includegraphics[height=3.4cm] {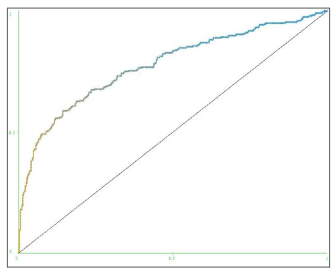}
\caption{Model based on meta and audio features}
\end{subfigure}
\begin{subfigure}[h]{0.32\textwidth} \centering
\includegraphics[height=3.4cm] {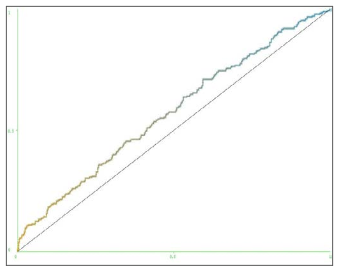}
\caption{Model based on audio features}
\end{subfigure}
\begin{subfigure}[h]{0.32\textwidth} \centering
\includegraphics[height=3.4cm] 
{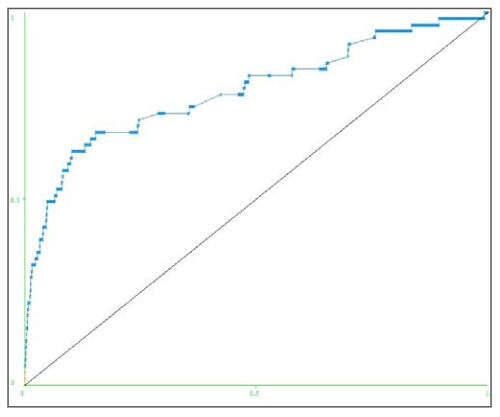}
\caption{Model based on early adopters}
\end{subfigure}
\caption{Receiver operating curves (ROC) for logistic regression models. True positive rate is displayed on the y-axis, and the false negative rate on the x-axis. }
\label{fig:roc}
\vspace{-.5cm}
\end{figure*}

Listening behaviour data for three Last.FM groups, `Consistently New Dance and Electronic' (2,116 members), `Electronic Music' (3,747 members) and `Belgium' (1,859 members), was parsed using the Last.FM API. Listening data was retrieved for a period of 6 months, from 16/04/13 until 16/11/13. This data was thoroughly cleaned to compensate for irregular spelling and abbreviations, e.g. 'featured' versus 'ft'. The resulting cleaned dataset consists of 854,060 listening records (time, date, user, song, artist). This dataset was further processed to contains a row for each song at each week (instance), and a column (feature) for each user indicating the number of times they listened to that particular song. A feature called `predictive' was set to 1 if the song was not yet a hit now, but will be in the future.

\vspace{-.4cm}
\subsection{Audio features}
A total of 140 audio features, including those that capture a temporal aspect, were extracted through the EchoNest API, as described by \cite{herremans2014dance}. Among the EchoNest features are a number of meta-features, such as danceability and hotness. As the EchoNest documentation does not provide a precise definition of these features, we have decided to create a second dataset whereby these features are removed, so as to be able to properly analyze the effect of acoustic properties only. In the next section, we refer to the dataset that includes all features as the `Meta feature' dataset and the reduced dataset as the `Audio feature' dataset.

\vspace{-.3cm}

\section{Results}
\label{sec:results}

\vspace{-.1cm}
\subsection{Audio feature-models}

Five classification models were built in Weka in order to classify top 20 hits from non-hits, as inspired by \cite{herremans2013composer, herremans2015composerbook}: RIPPER ruleset (RR), Logistic Regression (LR), Support Vector Machines (SVM) and Naive Bayes (NB). The AUC values (area under the receiver operating curve) are displayed in Table~\ref{tab:results}. 

A fairly high AUC value is obtained (0.77) when using all of the features. However, to be fully independent from The EchoNest's precalculated features such as `hotness', which might already include hit data, the audio feature dataset should be used. Models based on these features obtain a much lower maximum AUC (0.64), with the most accurate model being the logistic regression. The receiver operating curves of the logistic regression model for all datasets is shown in Figure~\ref{fig:roc}. 

\begin{table}[h!]
\caption{AUC results with 10-fold cross-validation.}
\label{tab:results}
\footnotesize
\begin{tabular}{lccccc}
\toprule
& C4.5 & RR & NB & LR & SVM  \\
\midrule
Meta features & 0.73 & 0.72 & 0.75 & \textbf{0.77} & 0.70 \\
Audio features & 0.60 & 0.61 & 0.60 & \textbf{0.64} & 0.59 \\
Early Adopters & 0.50 & 0.51 & 0.70 & \textbf{0.79} & 0.50 \\
\bottomrule
\end{tabular}
\end{table}

\subsection{Early adopters-based models}

When building models based on social media listening behaviour, overall higher AUC values are reached. The AUC of each of the models is displayed in Table~\ref{tab:results}. The best performing model, reaching an AUC of 0.79 is logistic regression. Currently, no parameter tuning for the SVM was performed, e.g. through grid-search, or by trying different kernels. This could potentially drastically improve the performance of the model and should be explored in future research. The current values, however, already clearly show that listening data from Last.FM can help us predict top 20 hits.

\vspace{-.4cm}

\section{Conclusions}

In this study, a large dataset of social listening behaviour gathered through the Last.FM API was created. The dataset also contained hit/non-hit information from the Belgian Dance charts Ultratop 50, and audio data gathered from EchoNest. Based on an early adopter model, the hit prediction models built in this research are able to predict upcoming top 20 hit with an AUC value of 0.79. 

In future research, it would be interesting to perform parameter tuning of the models and look at an embedded model that combines both audio and social listening data to further enhance hit prediction accuracy. In order to advance this field, it would also be beneficial to have an open dataset with fixed hit definitions for benchmarking.

% \begin{figure}
%  \includegraphics[width=0.48\textwidth]{c45audio.png}
% \end{figure}

% \begin{verbatim}
% If (loudness <= 0.573658) => klasse=A (129.0/55.0)
% If (duration <= 0.501363) and (duration >= 0.370967) => klasse=A
% (137.0/63.0)
% Other => klasse=B (378.0/127.0)
% \end{verbatim}

\footnotesize
\bibliography{paper}

\end{document}